\documentclass[
]{ceurart}

\usepackage{listings}
\usepackage{amsmath}
\usepackage{graphicx}
\begin{document}

\copyrightyear{2023}
\copyrightclause{Copyright for this paper by its authors. Use permitted under Creative Commons License Attribution 4.0 International (CC BY 4.0).}

\conference{2nd Edition of EvalRS: a Rounded Evaluation of Recommender Systems, August 6-August 10, 2023, Long Beach, CA, USA}

\title{Metric@CustomerN: Evaluating Metrics at a Customer Level in E-Commerce}
\subtitle{Discussion Paper}

\author[1]{Mayank Singh}[%
email=mayanksingh@grubhub.com,
orcid=0009-0005-5279-5049,
]
\address[1]{Grubhub, New York, USA}

\author[1]{Emily Ray}[%
email=eray1@grubhub.com,
]
\author[1]{Marc Ferradou}[%
email=mferradou@grubhub.com,
orcid=0000-0002-9693-4996
]
\author[1]{Andrea Barraza-Urbina}[%
email=abarraza@grubhub.com,
orcid=0009-0001-0342-5181
]

\begin{abstract}
Accuracy measures such as Recall, Precision, and Hit Rate have been a standard way of evaluating Recommendation Systems. The assumption is to use a fixed Top-N to represent them. We propose that median impressions viewed from historical sessions per diner be used as a personalized value for N. We present preliminary exploratory results and list future steps to improve upon and evaluate the efficacy of these personalized metrics.
\end{abstract}


\maketitle

\section{Introduction}

Recommender Systems (RS) are ubiquitous in e-commerce, from serving relevant ads to customers to helping them pick their favorite food. We have been evaluating these RS in the same manner for more than a decade using Metric@N\cite{Herlocker, Cremonesi}; e.g. \textit{Recall@N} and N takes a numeric value such as: {1, 5, 100}. Evaluating the performance of the system using a static N for all customers misses important nuances in their behavior on the platform\cite{Hosey}. Customer A might only look at the first 5 results on average but Customer B's average is 25. The prevailing industry assumption is that displaying “best” results on top is the optimal solution for an online RS, but this may not hold universally\cite{Hosey}. Some customers might not click on the first result even if it is the most relevant, because they want to “explore” additional results before making a decision. In line with the goal of EvalRS2023\cite{Bianchi}; we propose calculating a personalized evaluation metric at \textit{CustomerN} instead of a static $N$ termed: \textit{Metric@CustomerN}. One way of calculating \textit{CustomerN} is to take the median of maximum impression ranks scrolled to in past sessions on the platform.

\section{Related Work}

To the best of our knowledge, there are no other texts that discuss the use of a dynamic \textit{N} value while calculating accuracy metrics to evaluate a RS. Giobergia\cite{Giobergia} introduces "variance agreement" to account for different user interests on a music streaming platform. Chia et al.\cite{Chia} introduced \textit{RecList}, to standardize behavioral metrics testing, and also introduce data slice-based evaluation. Similarly, Ekstrand et al.\cite{Ekstrand} break down users by demographic groups to understand if users from different groups obtain the same utility from RS. Kaminskas et al.\cite{Kaminskas} expand beyond accuracy measures and study the non-accuracy measures such as Diversity, Serendipity, Novelty, and Coverage and discuss their calculation. Sun\cite{Sun} and Verachtert et al.\cite{Verachtert} detail the importance of observing a global timeline while evaluating recommender models. Using impression data in RS improved the relevance of recommended results in \cite{Perez, Aharon}, we propose incorporating impression data in RS evaluation as well.

\section{Metric@CustomerN}

The methodology to calculate \textit{Recall@CustomerN}\footnote{Recall is used as an example metric for representation. The same steps can be followed to calculate other similar metrics: Precision, Accuracy, Hit Rate, NDCG, etc.} is detailed in the steps below:
\begin{enumerate}
    \item For a customer $C_i$ in a set of customers $S$ we capture the max impression rank, $R_{ij}$, scrolled-to in each session $j$.
    \item We calculate the median impression position for a customer for sessions browsed in the last $X$ days:
    \begin{gather}
        N_i = \text{median}(R_{ij}), \quad \ i \in \{1,S\}, \quad j \in \{1, p_i\}
    \end{gather}
    where $p_i$ denotes the number of sessions browsed by customer $C_i$ and $X$ is decided based on platform and analysis goals.
    \item Now we can calculate the recall value for each customer denoted by: \textit{Recall@$N_i$}.
    \item For a summarized view of how the recommendation algorithm performs, we use average $Recall@N_i$ for all customers on the platform:
    \begin{gather}
        \frac{1}{S} \sum_{i=1}^{S} Recall@N_i, \quad \ i \in \{1, S\}
    \end{gather}
\end{enumerate}

\section{Preliminary Analysis}

\begin{figure}[h]
\centering
\includegraphics[width=\textwidth]{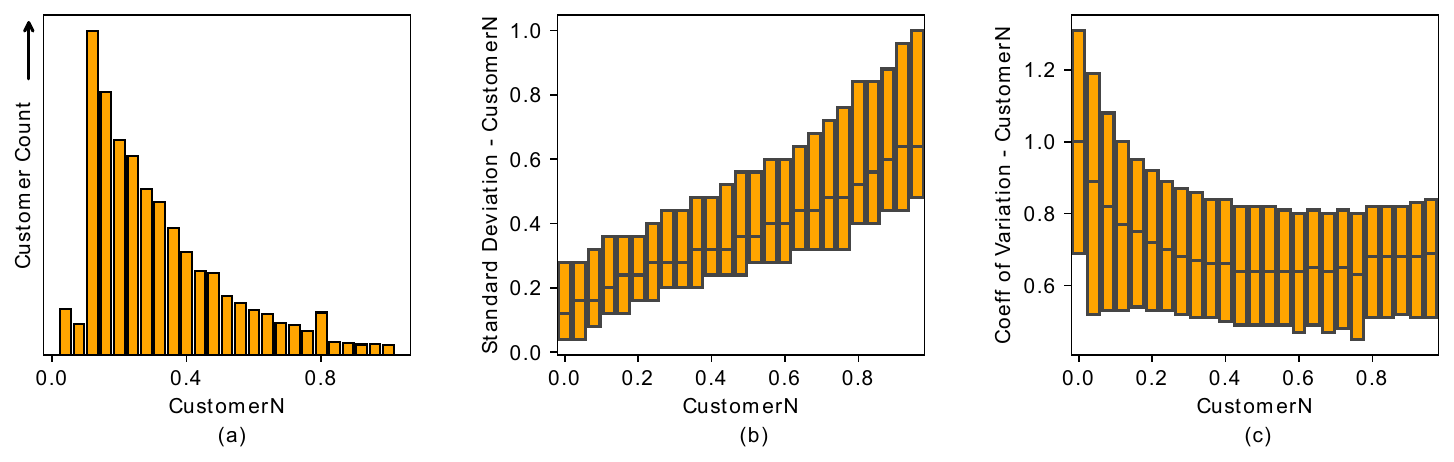}

\caption{CustomerN variability across customers with 3+ sessions in the last 90 days on Grubhub. All Axes have been normalized by max CustomerN. Bars represent the inter-quartile range of y-axis values in (b) and (c)}
\label{fig:combined}
\end{figure}

\textbf{Figure \ref{fig:combined}a} shows significant variation in CustomerN, supporting the need for diner-specific N. 

In \textbf{Figure \ref{fig:combined}b} and \textbf{Figure \ref{fig:combined}c} we observe that as the median impressions go up for a customer, so does the variance of their impressions viewed across sessions. Additionally, the CV value is higher for smaller \textit{CustomerN} and stabilizes for diners with higher median impression views.

\section{Discussion}

Based on the findings from \cite{Ekstrand, Ji, Kaminskas} and other research on improving the evaluation of RS, it is clear that we are trying to understand how to better explain variability in customer behavior on e-commerce platforms. As a future undertaking we would first compare the performance of popular RS algorithms on public\cite{Wu, Eide, PerezMaurera} and proprietary datasets using \textit{Metric@CustomerN}. Secondly, using median impressions viewed across all sessions as \textit{CustomerN} has its limitations because it cannot account for additional variability within the same customer's sessions as seen in Figure \ref{fig:combined}. So we would like to segment customer sessions based on their mindset per session using same-session variables, historical activity, demographics, and geographical variables as detailed in \cite{Cheng, Garcia} and subsequently calculate \textit{CustomerN} as median impressions viewed at the Customer-Segment level. Lastly, we will monitor long-term KPIs to validate if improved \textit{Metric@CustomerN} correlates with customer satisfaction and lifetime value.

\section{Conclusion}
Recent research \cite{Perez, Aharon} has shown us that it’s extremely valuable to incorporate customer impression data into an RS. Similarly, we propose using impression data to enhance the effectiveness of accuracy-based metrics. In our opinion, this approach has merit and warrants additional work to understand the implication of developing personalized calculations like \textit{Metrics@CustomerN} for RS evaluation. The preliminary analysis we did points to the existing variability in customer behavior and to a need for a customer-centric evaluation of accuracy metrics. The methodology described in this paper is just the first step toward building a more personalized evaluation outlook for RS, we look forward to testing it out at EvalRS2023\cite{Bianchi}.

\section*{Acknowledgments}

We would like to thank Ruonan Ding for laying the foundation for this work and Fan Gong for their advice and feedback on the paper.

\bibliography{metricsbib.bib}

\end{document}